\documentclass[10pt,aps,prl,twocolumn,floatfix,superscriptaddress]{revtex4}

\usepackage{subfigure}
\usepackage{epsfig}
\usepackage{dcolumn}
\usepackage{amsmath}
\usepackage{graphicx}
\usepackage{rotating}
\usepackage{multirow}
\usepackage{array}
\usepackage{amssymb}
\usepackage{enumerate}
\usepackage{epstopdf}
\usepackage{color}
\usepackage{float}
\usepackage{setspace}

\begin{document}

\title{Evolution of subway networks}

\author{Camille Roth}
\affiliation{CAMS (CNRS/EHESS) 190, avenue de France, F-75013 Paris, France\\}
\affiliation{Inst. Sys. Complexes Paris-Ile de France (ISC-PIF), 57-59 rue Lhomond, F-75005 Paris, France\\
  \texttt{\small roth@ehess.fr}}

\author{Soong Moon Kang}
\affiliation{Department of Management Science and Innovation\\
University College London (UCL), Gower Street, London WC1E 6BT, UK\\
 \texttt{\small smkang@ucl.ac.uk}}

\author{Michael Batty} \affiliation{Centre for Advanced Spatial
 Analysis (CASA)\\ University College London (UCL), 90 Tottenham Court Road, London W1N 6TR, UK\\
 \texttt{\small m.batty@ucl.ac.uk}}

\author{Marc Barthelemy}
\affiliation{CAMS (CNRS/EHESS) 190, avenue de France,
F-75013 Paris, France\\}
\affiliation{Institut de Physique Th\'eorique\\
CEA, IPhT, CNRS-URA 2306, F-91191 Gif-sur-Yvette, France\\
\texttt{\small marc.barthelemy@cea.fr}}

\begin{abstract}

  We study the temporal evolution of the structure of the world's
  largest subway networks in an exploratory manner. We show that,
  remarkably, all these networks converge to {a shape which shares
    similar generic features} despite their geographic and economic
  differences. This limiting shape is made of a core with branches
  radiating from it. For most of these networks, the average degree of
  a node (station) within the core has a value of order $2.5$ and the
  proportion of $k=2$ nodes in the core is larger than $60\%$. The
  number of branches scales roughly as the square root of the number
  of stations, the current proportion of branches represents about
  half of the total number of stations, and the average diameter of
  branches is about twice the average radial extension of the
  core. Spatial measures such as the number of stations at a given
  distance to the barycenter display a first regime which grows as
  $r^2$ followed by another regime with different exponents, and
  eventually saturates. These results -- difficult to interpret in the
  framework of fractal geometry -- confirm and yield a natural
  explanation in the geometric picture of this core and their
  branches: the first regime corresponds to a uniform core, while the
  second regime is controlled by the interstation spacing on
  branches. The apparent convergence towards a unique network shape in
  the temporal limit suggests the existence of dominant, universal
  mechanisms governing the evolution of these structures.

\keywords{Evolution of networks \and Urban transportation \and Spatial networks 
\and Core and branch geometry}
\end{abstract}

\maketitle

\section{Introduction}
\label{intro}

\begin{figure}[ht!]
\vspace{1em}
\centering
\begin{tabular}{c}
\includegraphics[width=0.7\linewidth]{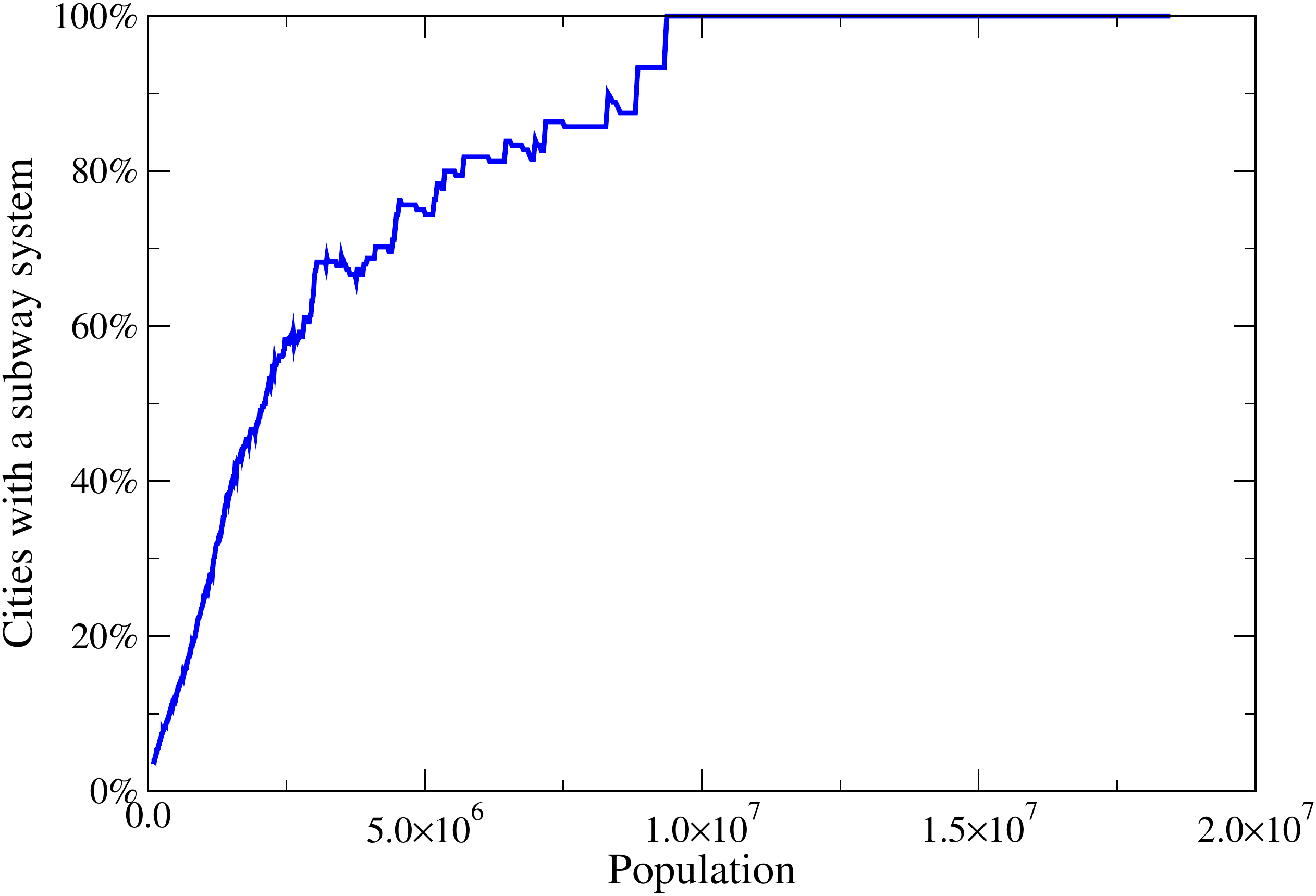}
\end{tabular}
 \caption{Percentage of cities with a subway system versus the
   population (data from the UN \cite{UNdata}).} \label{fig:percent}
\end{figure}

Transportation systems, especially mass transit, are an important
component in cities and their expansion. In a world where more than
$50\%$ of the population lives in urban areas \cite{UN}, and where
individual transportation increases in cost as cities grow larger,
mass transit and in particular, subway networks, are central to the
evolution of cities, their spatial organization
\cite{Hanson:2004,Batty:book,Malecki:2011} and dynamical processes
occurring in them \cite{Bettencourt:2007,Balcan:2009}. The percentage
$s(P)$ of cities with a subway system versus their population size $P$ is
shown in Fig.~\ref{fig:percent} (the data were obtained for cities
with population larger than $10^5$ \cite{UNdata}) which confirms that
the larger a city, the more likely it is to have some form of mass
transit system (see also \cite{Daganzo:2009}).
Approximately $25\%$ of the cities of more than one million
individuals have a subway system, $50\%$ of those of more than two millions,
and all those above $10$ millions have a subway system (as an
indication, an exponential fit of the plot in Fig.~\ref{fig:percent}
gives $s(P)=1-\exp(-P/P_0)$ where the typical population $P_0$ is of
order 3 millions).

For some cities, subway systems have existed for more than a
century. Fascination with the apparent diversity of their structure
has led to many studies and to particular abstractions of their
representation in the design of idealized transit maps
\cite{Ovenden:2003}, and although these might appear to be 
planned in some centralized manner, it is our contention here 
that subway systems like many other features of city systems evolve 
and self-organize themselves as the product of a stream of rational 
but usually uncoordinated decisions taking place through time.

Generally speaking, subway systems have been developed to improve
movement in urban areas and to reduce congestion. The early history of
subways is sometimes connected to large scale planning, for instance
with the need to bring population from a growing periphery to the
center where traditionally production and exchange usually take
place. More broadly, it might seem that subway systems are engineered
systems and intentionally structured in a core/periphery shape with
their self-organization thus playing only a very minor role. This
actually would be true if these subway systems were planned from their
beginning to their current shape, but this is not the case for most
networks. Their shape results from multiple actions, from planning
within a time limited horizon, set within the wider context of the
evolution of the spatial distribution of population and related
economic activities. We thus conjecture that subway networks actually
result from a superimposition of many actions, both at a central level
with planning and at a smaller scale with the reorganization and
regeneration of economic activity and the growth of residential
populations. In this perspective, subway systems are self-organizing
systems, driven by the same mechanisms and responding to various
geographical constraints and historical paths. This self-organized
view leads to the idea that --- beside local peculiarities due to the
history and topography of the particular system --- the topology of
world subway networks display general, universal features, within the
limits of the physical geometry and cultural context in which their
growth takes place.

The detection and characterization of these features require us to
understand the evolution of these spatial structures. Indeed, subway
networks are spatial \cite{Reggiani,Barthelemy:2011} in the sense that
they form a graph where stations are the nodes and links represent
rail connections. We now understand quite well how to characterize a
spatial network but we still lack tools for studying their temporal
evolution. The present article tackles this problem, proposing various
measures for these time dependent, spatial networks.

\begin{figure}[ht!]
\centering
\begin{tabular}{c}
\includegraphics[width=1\linewidth]{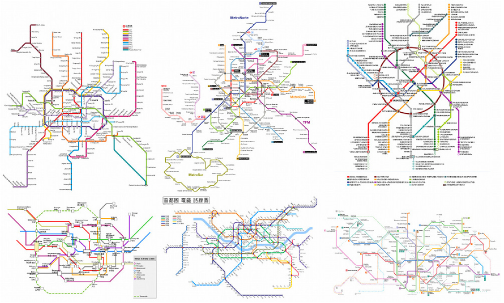}
\end{tabular}
 \caption{A sample of large subway networks in large urban areas, all
    displaying a core and branches structure. From left to right and
    top to bottom: Shanghai, Madrid, Moscow, Tokyo, Seoul, Barcelona
    (Figures from Wikimedia Commons
    \cite{figures}).}\label{fig:examples}
\end{figure}

Here we focus on the largest networks in major world cities and thus
ignore currently developing, smaller networks in many medium-sized
cities. We thus consider most of the largest metro networks (with at
least one hundred stations) which exist in major world cities. These
are: Barcelona, Beijing, Berlin, Chicago, London, Madrid, Mexico, Moscow, New York City
(NYC), Osaka, Paris, Seoul, Shanghai, and Tokyo, for which we show a sample in Fig.~\ref{fig:examples}.
Additionally, we focus on urban subway systems and do not
consider longer-distance heavy and light-rail commuting systems in urban areas, such
as RER (R\'eseau Express R\'egional) in Paris or overground NetworkRail in London.

Static properties of transportation networks have been studied for many
years \cite{Haggett:1969} and in particular simple connectivity
properties were studied in \cite{Bon:1979} while fractal aspects were
considered in \cite{Benguigui:1991}. With the recent availability of
new data, studies of transportation systems have accelerated
\cite{Barthelemy:2011} and this is particularly so for subway systems
\cite{Latora:2001,Seaton:2004,Sien:2005,Gattuso:2005,Fisk:2006,Lee:2008,Ferber:2009,Derrible:2010b,Derrible:2010a}.
These studies have revealed some significant similarities between
different networks, despite differences in their historical
development and in the cultures and economies in which they have been
developed.  In particular, their average shortest path seems to scale
with the square root of the number of stations and the average
clustering coefficient is large, consistent with general results
associated with two-dimensional spatial networks (see
\cite{Barthelemy:2011}). In \cite{Sien:2005}, a strong correlation
between the number of stations (for bus and tramway systems) and
population size has been observed for $22$ Polish cities, but such
correlation are not observed at the world level (for all public
transportation modes \cite{Ferber:2009}).

Our empirical analysis of the evolution of these transportation
networks is in line with approaches developed in the 1970's (see
\cite{Xie:2009} and references therein) but we take advantage here of
recent progress made in the understanding of spatial networks in
general and new historical data sources which provide us with detailed
chronologies of how these networks have developed.

\subsection{Data}
\label{sec:1.1}

The network topologies at various points in time were built using two
main data sources. First, current network maps as for 2009 were used to
define lines for each network, and then to define line-based
topologies, \hbox{i.e.} which station(s) follow(s) which other
station(s) on each line.  This information was then combined with
opening dates for lines and stations. This second type of data has
been gathered from Wikipedia \cite{Wikipedia}: for most networks,
there is one page per station with various information, including the
first date of operation, the precise location and address, number of
passengers, etc. The network building process for a given year $t$ is
then as follows. The list of open lines at year $t$ is first
established. For each open line, open stations at year $t$ are listed
and connections are created between contiguous stations according to
the network topology. A station which is not open at year $t$ on the
given line, even if it is already open on a different line, is
evidently discarded for the construction of the line.  Eventually,
those independent line topologies are gathered into the subway graph
corresponding to year $t$. Note that we used 2009 topologies as it was
relatively difficult to find and process network maps for all
these networks for each year of their existence.  As a result, topologies
for any given year before 2009 may overlook topology features
pertaining to station or line closures: for instance, a station which
existed between 1900 and 1940 and which remained closed until now will
not appear in any of our network datasets (such is the case for the British Museum Tube station).  We suggest however that
the effect of this bias is limited: on one hand, generally few
stations undergo closure in the course of the network evolution; on
the other hand, these stations are rarely hubs, most often
intermediary stations (of degree two, \hbox{i.e.} connected to two stations), thus their non-inclusion bears
little topological impact.

\section{Exploring static properties}
\label{sec:2}

The main characteristics of the networks we have chosen are shown in
Table~\ref{table:generalities} where we first observe that the number of different lines appears
to increase incrementally with the number of stations and that on
average for these world networks, there are approximately $18$
stations per line. Also, the mean interstation distance is on
average $\overline{\ell_1}\approx 1$km with Beijing and Moscow showing the longest
ones ($1.79$kms and $1.67$kms, respectively) and Paris displaying the shortest one ($570$ meters),
a diversity which finds its origin in the different historical paths
of these networks.
\begin{table*}[ht!]
\begin{tabular}{lccccccccc}
City & $P$ (millions) &$N_L$ & $N$ & $\overline{\ell_1}$ & $\ell_T$& $\ell_T/\ell_T^{reg}$ & $\beta$ $(\%)$\\
&&& &\footnotesize(kms)&\footnotesize(kms)&&\\
\hline
Beijing & $19.6$ & $9$ & $104$ & $1.79$ & $204$ & $0.14$ & $39$\\
Tokyo & $12.6$ & $13$ & $217$  & $1.06$& $279$& $0.13$ &$43$\\
Seoul & $10.5$ & $9$ & $392$ & $1.39$ &$609$ & $0.39$ & $38$\\
Paris & $9.6$ & $16$ & $299$ & $0.57$ & $205$ & $0.18$ &$38$\\
Mexico City & $8.8$ & $11$ & $147$ & $1.04$ &$170$& $0.15$&$39$\\
New York City\quad\hspace{.1em}  & $8.4$ & $24$ & $433$ & $0.78$ & $373$& $0.12$ &$36$ \\
Chicago & $8.3$ & $11$ & $141$ & $1.18$ & $176$ & $0.08$ & $71$\\
London  & $8.2$ & $11$ & $266$ & $1.29$ &$397$& $0.20$ &$47$\\
Shanghai & $6.9$ & $11$ & $148$ & $1.47$ &$233$& $0.21$ &$61$\\
Moscow & $5.5$ & $12$ &  $134$ & $1.67$ &$260$& $0.16$ &$71$\\
Berlin & $3.4$ & $10$ & $170$ & $0.77$ &$141$& $0.30$ &$60$\\
Madrid & $3.2$ & $13$ & $209$ & $0.90$ &$215$& $0.42$ &$46$\\
Osaka & $2.6$ & $9$ & $108$ & $1.12$ &$137$& $0.88$ &$43$\\
Barcelona & $1.6$ & $11$ & $128$ & $0.72$ &$103$& $0.32$ &$38$
\end{tabular}
\caption{ List of various indicators (for the
  year $2009$) for the major subway networks considered in this
  study (and sorted according to their metro population). $P$ is the
  metropolitan area population (for 2009). $N_L$ is
  the number of lines, $N$ the number of physical stations, $\overline{\ell_1}$ is the average 
  inter-station distance, $\ell_T$ total route length,
  $\ell_T^{reg}$ the total route length for a regular graph with
  same average degree, area, and number of stations, and $\beta$ the final ratio between branch and core stations.}
\label{table:generalities}
\end{table*}
Other quantities such as the catchment area (the average number of
individuals served by one station) could be computed but should be used
with care: residential and economic activity density vary strongly
across space and back-of-the-envelop arguments should only serve as a
guide.  Generally speaking, many parameters such as the population
density, land use activity distribution, and traffic are important
drivers in the evolution of those networks, but we will focus in this
first study on the characterization of these networks in terms of
space and topology, independently of other socio-economical
considerations. A later extension of this research could examine these
physical and topological properties with respect to various
definitions of density which might include different activity types
and various combinations related to the traffic that they generate.

In order to get some initial insight into the topology of these
networks, one can first compare the total length $\ell_T$ of these
networks to the corresponding quantity computed for an almost regular
graph $\ell_T^{\rm reg}$ with same number of
stations, area, and average degree (the ``degree'' of a node is the number of its neighbors in a graph). For a random planar graph with small degree fluctuations
($k\approx \langle k\rangle$) and small fluctuations of the spatial distribution of nodes, we can
consider that the internode spacing is roughly constant and given by
$\ell_0\sim 1/\sqrt{\rho}$ where $\rho=N/A$ is the density of nodes
defined as the number of nodes over the total area comprising all the
nodes. The total length is then the number of edges $E=N\langle
k\rangle/2$ times $\ell_0$ which leads to \cite{Barthelemy:2011}
\begin{equation}\label{eqlat}
\ell_T^{\rm reg}\sim \frac{\langle k\rangle}{2}\sqrt{AN}
\end{equation}
In real applications, the determination of the quantity $A$ is a
difficult problem, but here we choose to use the metropolitan area as
given by the various data sources. As shown in the
Table~\ref{table:generalities}, the ratio $\ell_T/\ell_T^{\rm reg}$
varies from $0.08$ to $0.88$, has an average of order $0.29$ and
displays essentially three outliers. First, Osaka (and also Madrid and
Seoul) has a very large value indicating a highly reticulated
structure. In contrast, Chicago and NYC have a much smaller value
($\approx 0.1$) signaling a more heterogeneous structure which in both
these cases is probably due to their strong geographical constraints.

The total length and the comparison with a regular structure gives a
first hint about the structure of these networks but other
indicators are needed to get a more focused view. There exist many different indicators and variables that
describe these networks and their evolution. An important difficulty
thus lies in the choice of the many possible indicators and how to extract useful
information from them. In addition, the largest networks have a
relatively small number of stations (always smaller than $500$) which
implies that we cannot expect to extract useful information from the
probability distributions of various quantities as the results are
too noisy. We thus have to compute more globally structured indicators
which are, however, sensitive to the usually small temporal variations
associated with these networks. In the following, we will focus on a
certain number of these indicators, which we consider to be the most
informative at this point.

Finally, we will focus in this study on purely spatial and topological
properties: we will consider the evolution in space of these subway
networks and we will not consider any other parameters which might be
used to characterize urban growth. Our study is exploratory and thus a
first step towards the integration of the most important factors into
this research and despite its simplicity, in that we focus almost
entirely on geometrical attributes, we consider that the evolution of the
topology encodes many different factors and that its study can point
to some important general mechanisms governing the evolution of these
networks.

\section{Network Dynamics}
\label{sec:3}

In order to get an initial impression of the dynamics of these
networks, we first estimate the simplest indicator $v=dN/dt$ which
represents the number of new stations built per year. From the
instantaneous velocity, we can compute the average velocity over all
years. This average can however be misleading as there are many years
where no stations are built and thus we describe this by the fraction
of `inactivity' time $f$. We provide results for the networks
considered in Table~\ref{table:dynamic} from which some
interesting facts are revealed.  Note that it is clear that Shanghai and
Seoul are the most recent subway networks experiencing a rapid
expansion that has elevated them to amongst the largest networks in
the world. 

\begin{table}[ht!]
\begin{tabular}{lccccc}
City & $t_0$ &~\hspace{.5em}\quad& $\overline{v}$ & $\sigma_v$ & $f$\\
\hline
Beijing                & $1971$ && $3.3$ & $7.74$ & $79\%$\\
Tokyo                & $1927$ && $2.8$ & $5.47$ & $51\%$\\
Seoul                 & $1974$ && $11.2$ & $14.9$ & $20\%$ \\
Paris                  & $1900$&& $2.6$ & $5.1$ & $60\%$ \\
Mexico City            & $1969$ && $3.7$ & $5.9$ & $55\%$ \\
New York City\quad\hspace{.1em} & $1878$ && $3.3$ & $8.3$ & $68\%$\\
Chicago                & $1901$ && $1.9$ & $6.24$ & $71\%$\\
London             & $1863$ && $2.3$ & $3.8$ & $48\%$ \\
Shanghai           & $1995$ && $14.9$ & $20.2$ & $31\%$ \\
Moscow            & $1936$ && $1.7$ & $1.9$ & $43\%$ \\
Berlin   		    & $1901$ && $1.6$ & $3.3$ & $65\%$\\
Madrid              & $1919$ && $2.3$ & $4.6$ & $59\%$ \\
Osaka 		& $1934$ && $1.4$ & $4.1$ & $79\%$ \\
Barcelona	& $1914$ && $1.4$ & $4.8$ & $78\%$\\
\hline
\end{tabular}
\caption{ $t_0$ is the initial year considered here for the different subways
  networks. $\overline{v}$ is the average velocity (number of stations
  built per year), $\sigma_v$ is the standard deviation of $v$, and $f$ is
  the fraction of years of inactivity (no stations built). 
  }
\label{table:dynamic}
\end{table}

\begin{figure}
\centering
\begin{tabular}{ll}
\footnotesize(A)&\footnotesize(B)\\\\
\includegraphics[width=0.49\linewidth]{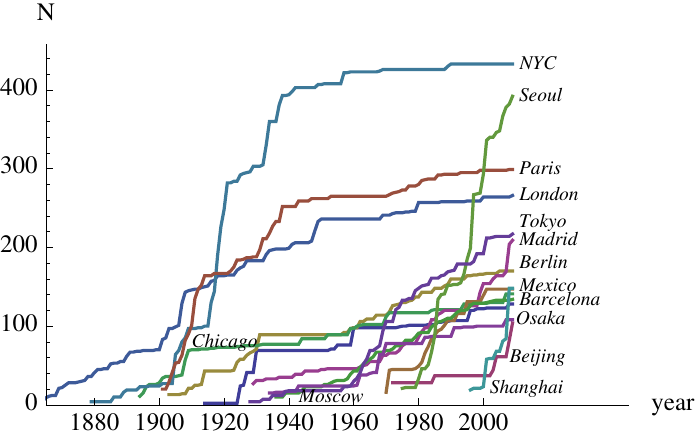}&
\includegraphics[width=0.49\linewidth]{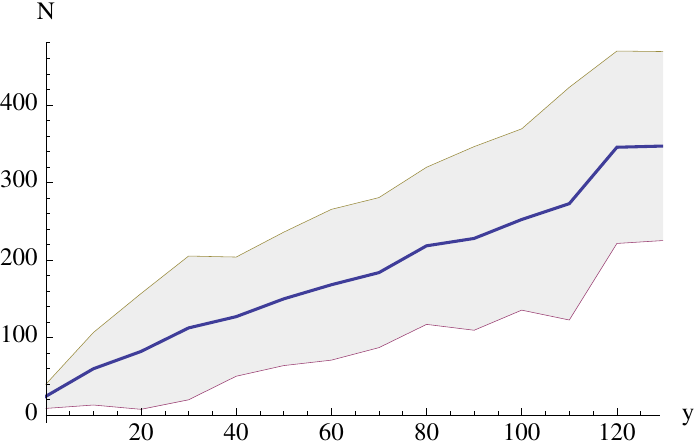}
\end{tabular}
\caption{(A) Evolution of the number of stations for various large
  world subway networks. (B) Evolution of the number of stations $y$
  years after creation, averaged over all networks (tubes mark the
  standard deviation across all networks). The linear shape indicates
  that the growth in terms of new stations from a decade to another
  goes to zero for all these networks, signaling the possible appearance of a
  stationary limit.}\label{fig:dyn1}
\end{figure}

For most of these networks the average velocity is in a small range
(typically $\overline{v}\in [1.4,3.7]$) except for Seoul and
Shanghai which are more recently developed networks. This is however
an average velocity and we observe that (i) for all networks, larger
velocities occur at earlier stages of the network and (ii) large
fluctuations occur from one year to another. Interestingly, the
fraction of inactivity time (\hbox{i.e.}  the time when no stations
are built) is similar for all these networks with an average of about
$58\%$. We also show in Fig.~\ref{fig:dyn1}(A), the time evolution for
each city of the number of stations, using an absolute time scale. In
particular, the size of the oldest networks seem to progressively
reach a plateau.

To make growth comparable across all networks, we introduce a second
graph on Fig.~\ref{fig:dyn1}(B) featuring the average, over all
networks, of the number of stations after a certain number of years
since network creation.  This average quantity exhibits a linear
increase which indicates convincingly that, overall, as these networks
become large, then for a few decades thereafter new stations represent
an increasingly small percentage of existing ones.  In other words, the
time evolution of all these networks is characterized by small
additions and not by sudden, abrupt changes with a large number of
stations added in a small time duration. This first result anticipates
the fact that these large networks may reach some kind of limiting
shape that we will characterize in the next section. This incremental
growth of subways might reflect socio-economical concerns and pressure
on the transportation networks such as diminishing return on
investments as noted by various authors (see for example
\cite{Naridi:1996} for US highways).

\begin{figure}
\centering
\begin{tabular}{c}
\includegraphics[width=0.8\linewidth]{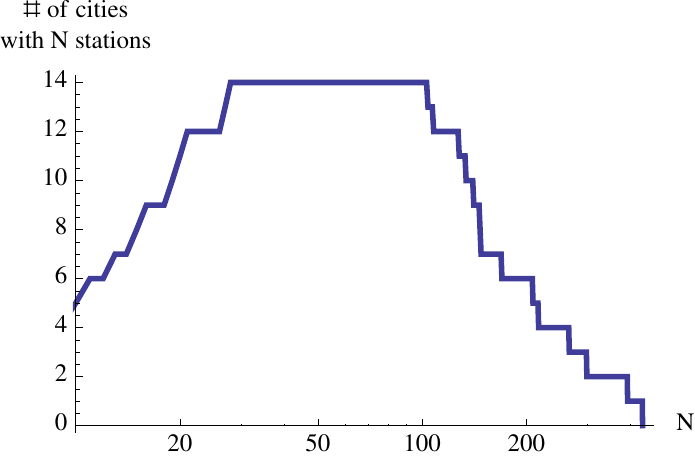}
\end{tabular}
\caption{ Number of cities with a given number of stations at a given
  time.}\label{fig:numcities} 
\end{figure}

Finally, when we study the evolution of various indicators versus the
number $N$ of station, an important point for our statistical analysis
is the number of subways with a given number $N$ at a given time
$t$. We show this quantity in Fig.~\ref{fig:numcities} 
and we can see that for $N\in[25,100]$ approximately this number is the largest 
(almost $15$ --- note that this figure is nonetheless too small to allow a 
discussion of the normality of the various quantities considered below). Unfortunately, 
for larger values of $N$ 
the number of cities is naturally smaller, and at this stage we cannot
give definitive answers but suggest some limits for large $N$.

\subsection{Characterization of the core and branches structure}
\label{sec:3.1}

The large subway networks considered here thus converge to a long time
limit where there is always an increasingly smaller percentage of new
stations added through time. The remarkable point that we will show
below is that all these networks, despite their geographical and
economical differences, converge to {a shape which exhibits several
  typical topological and spatial features}. Indeed, by inspection, we
observe that in most large urban areas, the network consists of a set
of stations delimited by a `ring' that constitute the `core'. From
this core, quasi-one dimensional branches grow and reach out to areas
of the city further and further from the core. In
Fig.~\ref{fig:examples}, we show a sample of these networks as they
currently exist. We note here that the ring, which is defined
topologically as the set of {core stations which are either at the
  junction of branches or on the shortest geodesic path connecting
  these junction stations,}
exists or not as a subway line. For instance, for Tokyo, there is a
such a circular line (called the Yamanote line), while for Paris the
topological ring does not correspond to a single line. It is also worth 
noting that in those systems where the core is harder to define such as NYC
where physical constraints are strongly manifest (the east and west rivers 
which bound Manhattan), a pseudo core is evident where a series of lines
coalesce to enable travelers to move around the core circumferentially.

More formally, branches are defined as the set of stations which are
iteratively built from a `tail' station, or a station of degree
1. New neighbors are added to a given branch as long as their degree
is 2 -- continuing the line, or 3 -- defining a fork. In this latter
case, the aggregative process continues \emph{if and only if} at least
one of the two possible new paths stemming from the fork is made up of
stations of degree 2 or less. Note that the core of a network with no
such fork is thus a $k$-core with $k=2$ \cite{Seidman:1983}.

\newcommand{\NB}{N_\text{B}} 
\newcommand{\NC}{N_\text{C}} 

\begin{figure}[h!]
\centering
\begin{tabular}{c}
\includegraphics[width=0.7\linewidth]{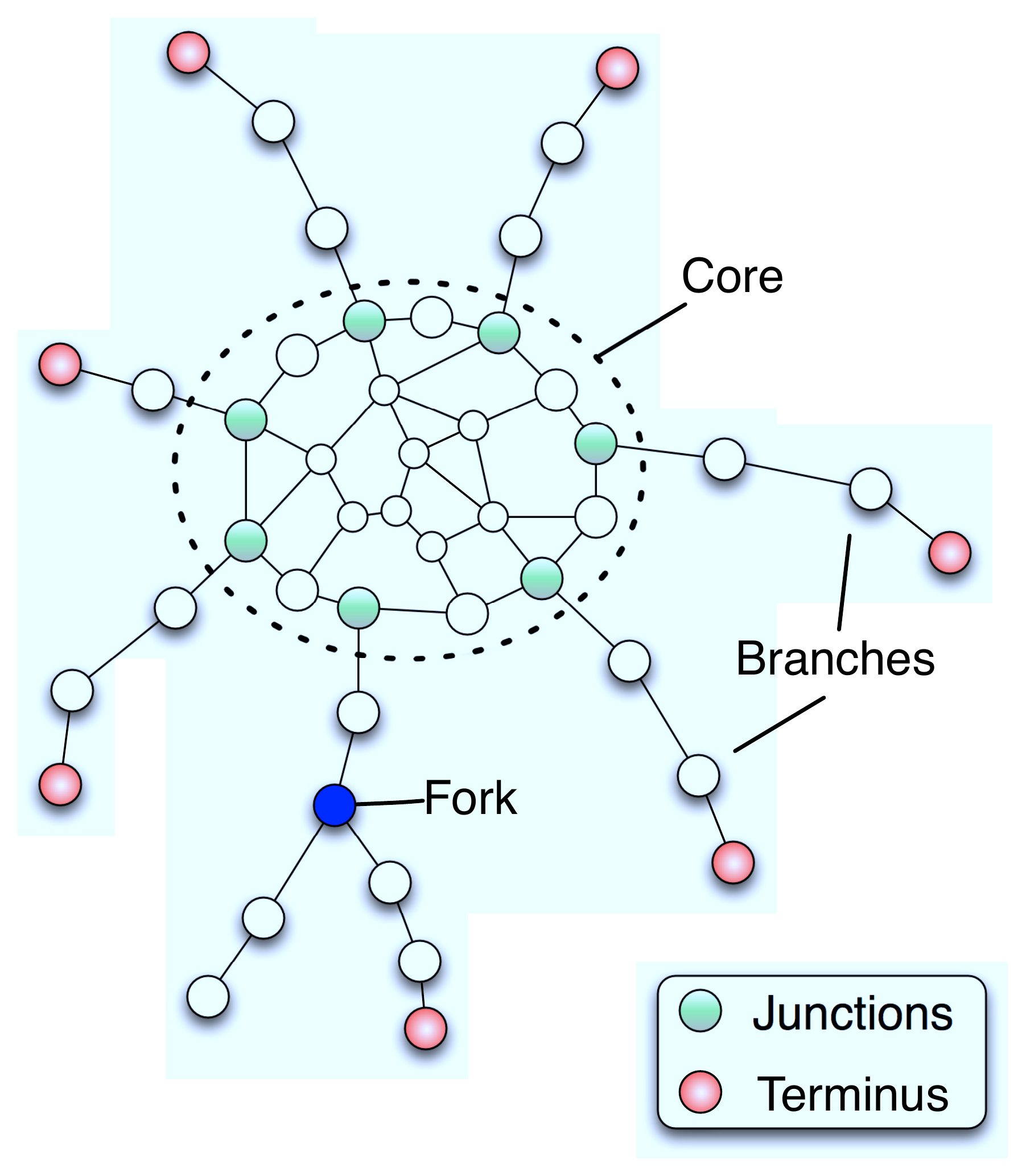}
\end{tabular}
\caption{Schematic structure of subway networks. A large `ring'
  encircles a core of stations. Branches radiate from the core and
  reach further areas of the urban system. The branches are
  essentially characterized by their size (parameter $\beta (t)$,
  Eq.~$2$), and their spatial extension (parameter $\eta (t)$ in
  Eq.~$3$). The core is characterized by its average degree ($\langle
  k_{\text core}\rangle (t)$ defined in Eq.~$4$) and fraction of nodes of degree 2 ($f_2$), its number of
  stations $N_C(t)$ and its size $r_C(t)$.}\label{fig:template}
\end{figure}

The general structure can schematically be represented as in
Fig.~\ref{fig:template}.  

We first characterize this branch and core structure with the 
parameter $\beta(t)$ defined as 
\begin{equation}
\beta(t)=\frac{\NB(t)}{\NB(t)+\NC(t)}
\label{eq:beta}
\end{equation}
where $\NB (t)$ and $\NC (t)$ respectively represent the number of
stations on branches and the number of stations in the core at time $t$.  

We can also characterize a little further the structure of
branches. Their topological properties are trivial and their
complexity resides in their spatial structure. We can then determine
the average distance (in kms) from the geographic barycenter of the city to all
core and branches stations, respectively: $\overline{D}_{\text{C}}(t)$
and $\overline{D}_{\text{B}}(t)$ (the barycenter is computed as the
center of mass of all stations, or in other words, the average
location of all the stations) This last distance provides
information about the spatial extension of the branches when we can
form the ratio $\eta (t)$
\begin{equation}
\eta (t)=\frac{\overline{D}_{\text{B}} (t)}{\overline{D}_{\text{C}}
  (t)}
\label{eq:eta}
\end{equation}
which gives a spatial measure of the amount of extension of the branches.

We also need information on the structure of the core. The core is a
planar (which is correct at a good accuracy for most networks)
spatial network and can be characterized by many parameters
\cite{Barthelemy:2011}. It is important to choose those which are not
simply related but ideally represent different aspects of the
network (such as those proposed in the form of various indicators, see
for example \cite{Haggett:1969,Xie:2007,Barthelemy:2011}). At each
time step $t$, we will characterize the core structure by the
following two parameters. The first parameter is simply the average
degree of the core which characterizes its `density'
\begin{equation}
\langle k_{\text{core}}\rangle(t)=\frac{2E_C(t)}{N_C(t)}
\label{eq:kcore}
\end{equation}
where $N_C(t)$ is the number of core nodes and $E_C(t)$ the number of
its edges. The average degree is connected to the standard index
$\gamma(t)=E_C(t)/(3N_C(t)-6)$ where the denominator is the maximum
number of links admissible for a planar network \cite{Haggett:1969}.

The average degree of the core contains a useful information about it,
and there are many other quantities (such as standard indices such as $\alpha$,
etc., see for example \cite{Haggett:1969}) which can give additional
information. We will use another simple quantity which describes in more detail the level of
interconnections in the core and which is given by the fraction $f_2$ of nodes in the
core with $k=2$. In the case of the well-interconnected system, this
fraction will tend to be small, while sparse cores with a few
interconnections will have a larger fraction of $k=2$ nodes.

Once we know this fraction $f_2$ of $k=2$ nodes in the core which
characterizes the level of interconnection and the parameter $\eta(t)$
which characterizes the relative spatial extension of branches, we
have key information on the intertwinement of both topological and
geographical features in such ``core/branch'' networks.

\subsection{Time evolution of $\beta$, $\langle k_{\text{core}}\rangle$, $f_2$ and
  $\eta$}
\label{sec:3.2}

The historical development of these networks is very different from
one city to another and representing the evolution of a specific
quantity versus time would probably not be particularly
meaningful. Similarly, city networks often experience significant
development in some particular years, while they experience
little or no evolution for the rest of the time. In order to be able
to compare the networks across time periods and cities, we propose
to study their evolution in terms of the number of stations $N$ that are
constructed.

\begin{figure}[ht!]
\centering
\vspace{1em}
\begin{tabular}{ll}
\footnotesize(A)&\\
&
\includegraphics[width=0.8\linewidth]{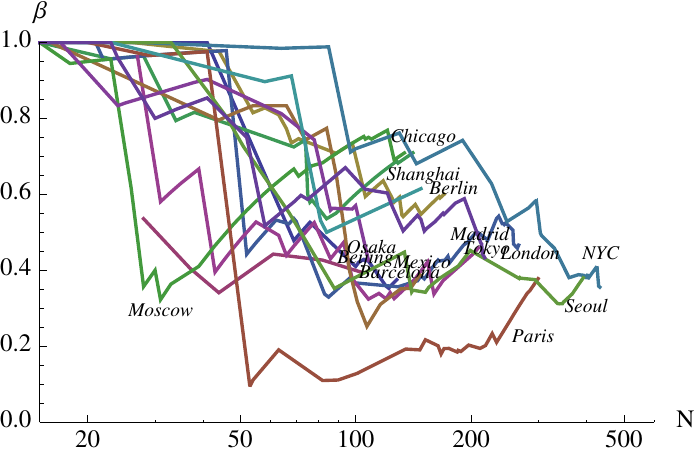}\\\\
\footnotesize(B)&\\
&
\includegraphics[width=0.8\linewidth]{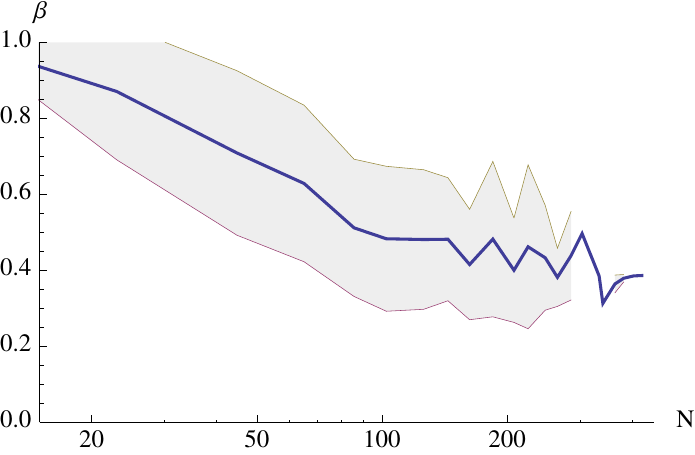}
\end{tabular}
 \caption{(A) Parameter $\beta$ as a function of the number of stations
   $N$ for the different world subways. (B) Same as (a) but averaged
   over 20 bins and showing the standard deviation.} \label{fig:beta}
\end{figure}

We first plot in Fig.~\ref{fig:beta}(A) the parameter $\beta$ as a
function of $N$ for the networks studied here. It is difficult to draw 
strong conclusions from this plot, but we can bin these data and
represent the average value of $\beta$ per bin and its dispersion as
well (Fig.~\ref{fig:beta}(B)). On this figure we may see that the
average value of $\beta$ seems to stabilize slowly to some value in $[0.35,0.55]$.

\begin{figure}[h!]
\vspace{1em}
\centering
\begin{tabular}{c}
\includegraphics[width=0.8\linewidth]{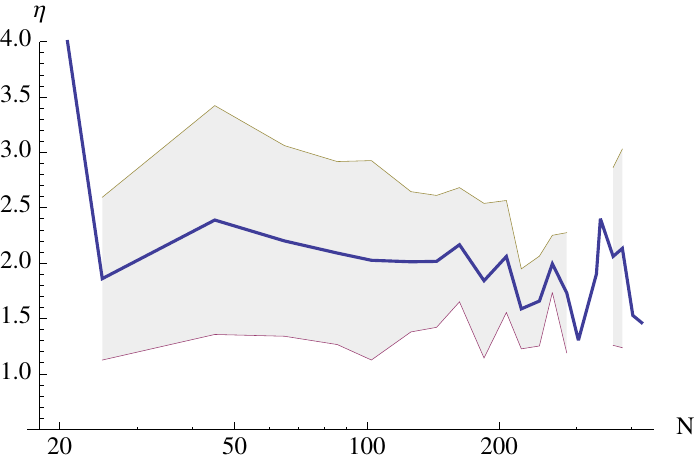}
\end{tabular}
 \caption{Evolution of the ratio $\eta$, which
    characterizes the spatial extension of branches relative to the
    core.}\label{fig:eta}
\end{figure}

It is also important to characterize the spatial importance of the
branches. The parameter $\eta$ gives a precious indication about their
extension and we show in Fig.~\ref{fig:eta} the evolution of this
parameter with $N$ (the data is binned).
This figure shows that in the interval where we have the largest
number of subways, the average value of $\eta$ is around $2$ with relatively 
large fluctuations which seem to decrease with $N$.

\begin{figure}[h!]
\vspace{1em}
\centering
\begin{tabular}{ll}
\footnotesize(A)&\\
&
\includegraphics[width=0.8\linewidth]{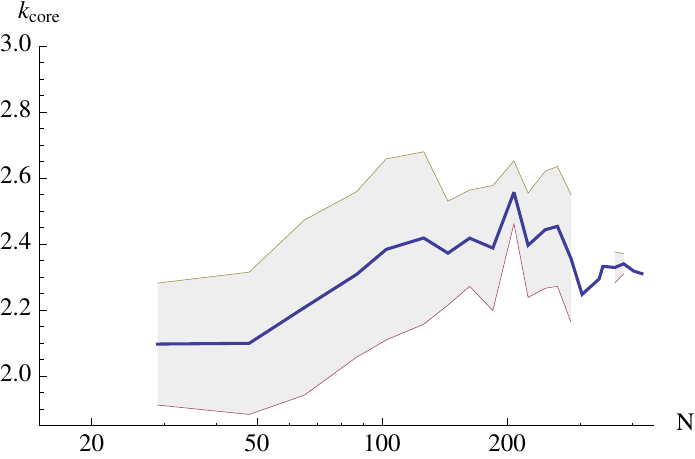}\\\\
\footnotesize(B)&\\
&
\includegraphics[width=0.8\linewidth]{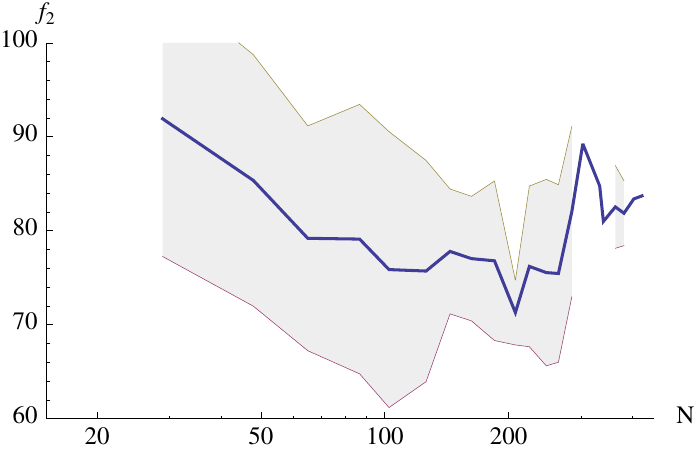}
\end{tabular}
 \caption{ (A) Average degree of the core $\langle
    k_{\text{core}}\rangle$ (Eq.~$4$) and its dispersion versus number of stations
    (averaged over $20$ bins). (B) Evolution of the 
      percentage $f_2$ of $k=2$ core nodes (averaged over $20$ bins).}\label{fig:evolution}
\end{figure}

The parameters $\beta$ and $\eta$ give an indication of the importance
of the core but do not say anything about its structure. A first
structural indication may be given by its average degree $\langle
k_{\text{core}}\rangle$ and by the percentage $f_2$ of nodes in the core
having a degree $k$ equal to $2$. In particular, these two quantities
shed light on how interconnections are created in the core.  We
display in Fig.~\ref{fig:evolution}(A) the average degree of the core
$\langle k_{\text{core}}\rangle$ which, even if there is a slow
increase with $N$, displays moderate variations around $2.4$
approximately. 

\begin{figure}[h!]
\vspace{1em}
\centering
\begin{tabular}{c}
\includegraphics[width=0.8\linewidth]{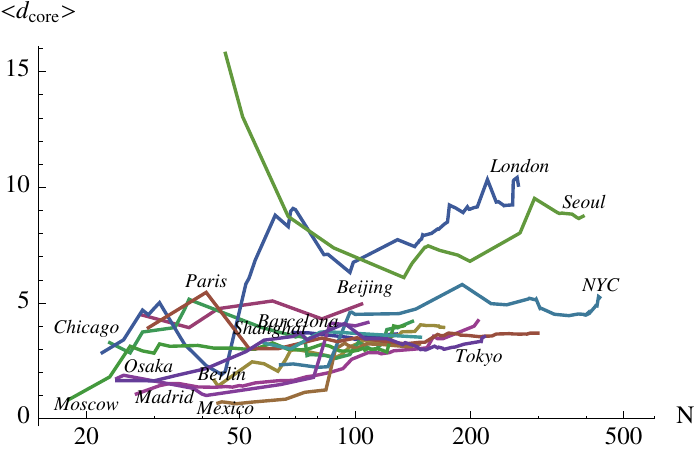}
\end{tabular}
 \caption{ Evolution of the mean distance to the barycenter (in kms) of core stations with the
   number of stations $N$.}\label{fig:evoldcore}
\end{figure}

This value is relatively small and indicates that the
fraction of connecting stations (\hbox{i.e.} with $k>2$) is also small
and means that most core stations belong to one single line with few
that actually allow connections. More precisely, we observe in
Fig.~\ref{fig:evolution}(B) that on average for subways with $N<100$
the fraction of interconnecting stations is increasing with $N$ --
which probably corresponds to some organization of the subway -- but
that for larger subways ($N>100$), the percentage $f_2$ is
increasing again, which probably corresponds to a densification
process without the creation of new interconnections. This
densification can indeed be confirmed as the diameter of the core (see
Fig.~\ref{fig:evoldcore}) seems to reach a plateau for most cities.


As noted above, the number of subways with large $N$ is smaller and
the statistics therefore less reliable. At this point and with this
statistical error in mind, we observe that the average value $\beta$
and its dispersion are decreasing with $N$ and it suggests that $\beta$
could converge to some `limiting' value $\beta_\infty\approx
45\%$. The same remarks also apply to $\eta$ and suggest a limiting
value of order $2$. Concerning the core, the dispersion of $\langle
k_{\text{core}}\rangle$ is always moderate and approximately
constant showing that the fluctuations among different networks are
also moderate. We observe a slow increase of $\langle
k_{\text{core}}\rangle$ pointing to a mild yet continuing
densification of the core, even after a long period of time. The
fraction of connecting stations has a more complex dynamics and seems
to decrease with $N$ for large networks. In these networks, there is
an obvious cost associated with the large value of $k>2$ and such a
decreasing fraction could be due to the fact that a small fraction is
enough to enable easy navigation in the network.

In summary, our results display non negligible fluctuations but
suggest that large subway networks may converge to a long time
limiting network largely independent of their historical and
geographical differences. So far, we can characterize the `shape' of
this long time limiting network with values of $\beta_{\infty}\approx
45\%$, $\eta_{\infty}\approx 2$, and a core made of approximately $80\%$ of
non connecting stations. It will be interesting to observe the future
evolution of these networks in order to confirm (or not) our current
results.

\subsection{Number of branches}

We now consider the number ${\cal N}_B$ of different 
branches. A naive argument would be that the number of branches is
actually proportional to the perimeter of the core structure. This
implicitly assumes that the distance between different branches is
constant. In turn, the perimeter should roughly scale as $\sqrt{N}$ as
the core is a relatively dense planar graph and contains a number of
nodes proportional to $N$. These assumptions thus leads to
\begin{equation}
{\cal N}_B\sim \sqrt{N}
\end{equation}
We display the number of branches versus the
number of stations $N$ for the various networks considered here.
\begin{figure}[h!]
\vspace{1em}
\centering
\includegraphics[width=0.7\linewidth]{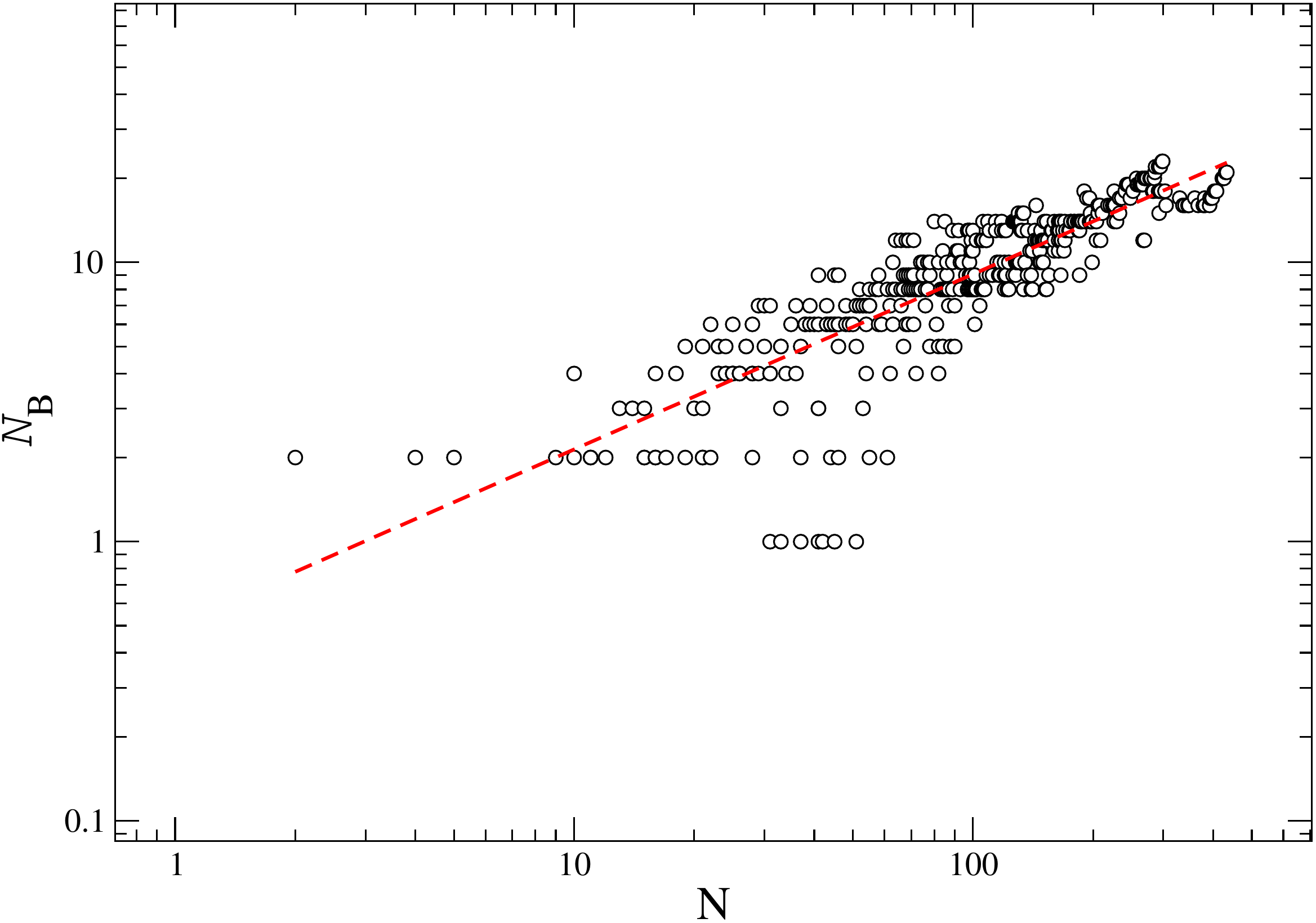}
\caption{ Loglog plot of the number of different branches versus the
  number of stations for the different subway networks considered
  here. The dashed line is a power law fit with exponent $\approx 0.6$.}\label{fig:NB}
\end{figure}
A power law fit of the data presented in figure~\ref{fig:NB} gives ${\cal
  N}_B\sim N^b$ with $b\approx 0.6$ ($r^2=0.85$) consistent with our
argument.

\subsection{Balance between the core density and the branch structure}
\label{sec:3.3}

\begin{figure}[h!]
\vspace{1em}
\centering
\includegraphics[width=0.7\linewidth]{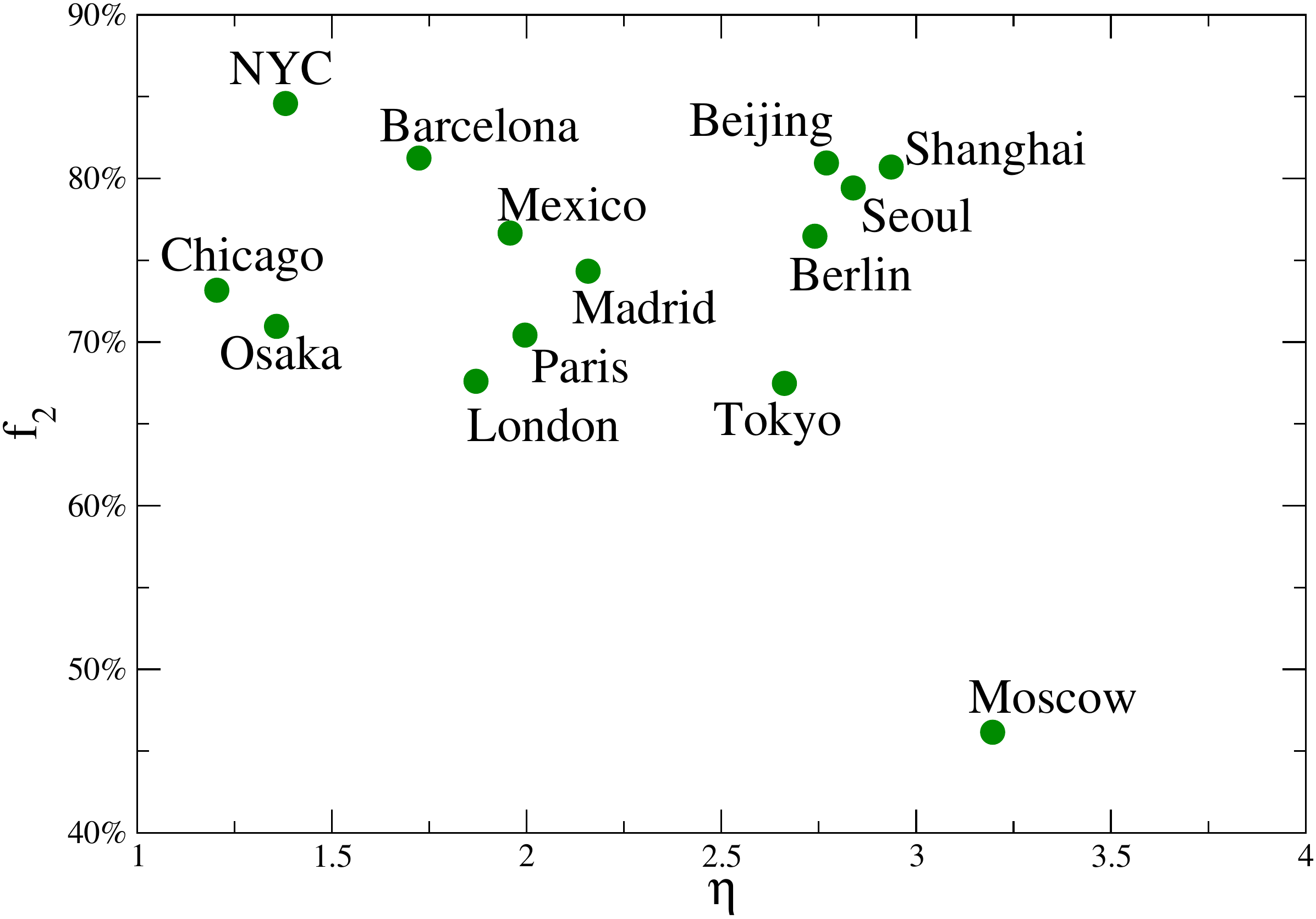}
\caption{ Relation between the spatial extension of branches and the
  degree of interconnection in the core. The $2009$ values for the
  percentage $f_2$ of $k=2$ core nodes and $\eta$ are plotted for
$12$ city subways.}\label{fig:maturity} 
\end{figure}

Even if it seems that the values of various indicators converge with
the size of the networks, we still have appreciable variations. For
example $\eta$ varies from $\approx 1$ to $\approx 3$ and exhibits
a relatively constant and not negligible relative dispersion. It is
thus important to understand the remaining differences between these
networks. To achieve this, we focus on the relation between $\eta$
which characterizes the spatial extension of the branches relative to
the core, and the percentage $f_2$ of $k=2$ nodes in the core which
indicates how well connected the core is. We focus on the `final'
values of these parameters obtained for $2009$ for the various
networks and we obtain the plot shown in Fig.~\ref{fig:maturity}.
From this figure, we first see that $(\eta,f_2)$ ranges from
$(\approx 1.4,\approx 85\%)$ for NYC up to $(\approx 3.3,\approx 45\%)$
for Moscow which is indeed a highly ramified network with a very dense
core.

Very roughly speaking, we first observe that for this set of the
largest subway systems in the world, the percentage $f_2$ is large
and above $60\%$ and relatively independent from $\eta$. At a finer
level, we observe from this figure that clusters of networks with
similar properties also emerge. The first cluster comprises Beijing,
Berlin, Shanghai, and Seoul which are remarkably close to each other:
$f_2$ is of order $80\%\pm 5\%$ and $\eta(t)\approx 2.84\pm
0.1$. This cluster corresponds thus to subway networks with a large
degree of ramification and a lower interconnection level in their
core. Not surprisingly, this cluster comprises rapidly evolving
networks such as Beijing and Shanghai for example. Another cluster
comprises London, Paris and Madrid with a smaller value of
$f_2\approx 70\%\pm5\%$ which might result from their denser city
center structure and a smaller value of $\eta\approx 2$. This other
cluster corresponds to denser networks, less ramified but with more
interconnections in the core. Finally we can identify another cluster
made of Chicago and Osaka with a small value of $\eta$ and a
relatively dense core (with $f_2\approx 70\%$).

\section{Spatial organization of the core and branches}
\label{sec:4}

Following earlier studies on the fractal aspects of subway networks
\cite{Benguigui:1991}, we can inspect the spatial subway organization
by considering the number of stations $N(r)$ at a distance less than
or equal to $r$, where the origin of distances is the barycenter of
all stations considered as points. Interestingly, the barycenter of
all stations is almost motionless, except in the case of NYC where the
barycenter moves from Manhattan to Queens and thus we will exclude NYC
from further study. Chicago is a similar case: the spatial
structure of the core is peculiar, mainly due to presence of the
lake which constrains the network from expanding in the other
directions. We will also exclude this network in this section. It
should however be noted here that both Chicago and NYC do follow the
image of core and branches but that the main difference with the other
networks is that the core of these networks has no clear spatial
meaning due to the geographical constraints (such as the presence of a
lake for Chicago and a particular land area shape for NYC).

For the year 2009, the limiting shape made of a core and branches
implies that there is an average distance $r_C$ which determines the
core. In practice, we can measure on the network the size $N_C$ of the
core and we then define $r_C$ such that $N(r=r_C)=N_C$ (which assumes
implicitly an isotropic core shape, which is the case for most
networks except for the excluded cases of Chicago and NYC).  For the various
cities, we can easily compute the function $N=N(r)$ from which we can
extract $r_C$ and we report the results in the Table~\ref{table:rc}.

\begin{table}[ht!]
\begin{tabular}{lcc}
\hline
City & $N_C$ &  $r_C$ (kms)\\
\hline
Beijing               & $63$   & $4.4$    \\
Tokyo                & $123$ &  $5.0$     \\
Seoul                 & $243$ &  $11.6$   \\
Mexico             &  $90$   &  $4.7$      \\
Shanghai           & $57$   &  $3.7$      \\
Moscow            & $39$   &   $5.9$       \\
London             & $142$  &   $7.3$ \\
Paris                  & $186$ &   $4.2$ \\
Madrid              & $113$  &   $4.4$ \\
Berlin   		    & $68$ &   $5.5$\\
Barcelona   		    & $57$ &   $3.5$\\
Osaka   		    & $46$ &   $3.6$\\
\hline
\end{tabular}
\caption{ For each city, we compute the number of
  stations in the core (for the year 2009) and from the numerical calculation of $N(r)$ we
  can estimate $r_C$ the size of the core (in kms) from $N(r=r_C)=N_C$.}
\label{table:rc}
\end{table}

\begin{figure}[h]
\centering
\begin{tabular}{lr}
\footnotesize(A)&\\
&
\includegraphics[width=0.7\linewidth]{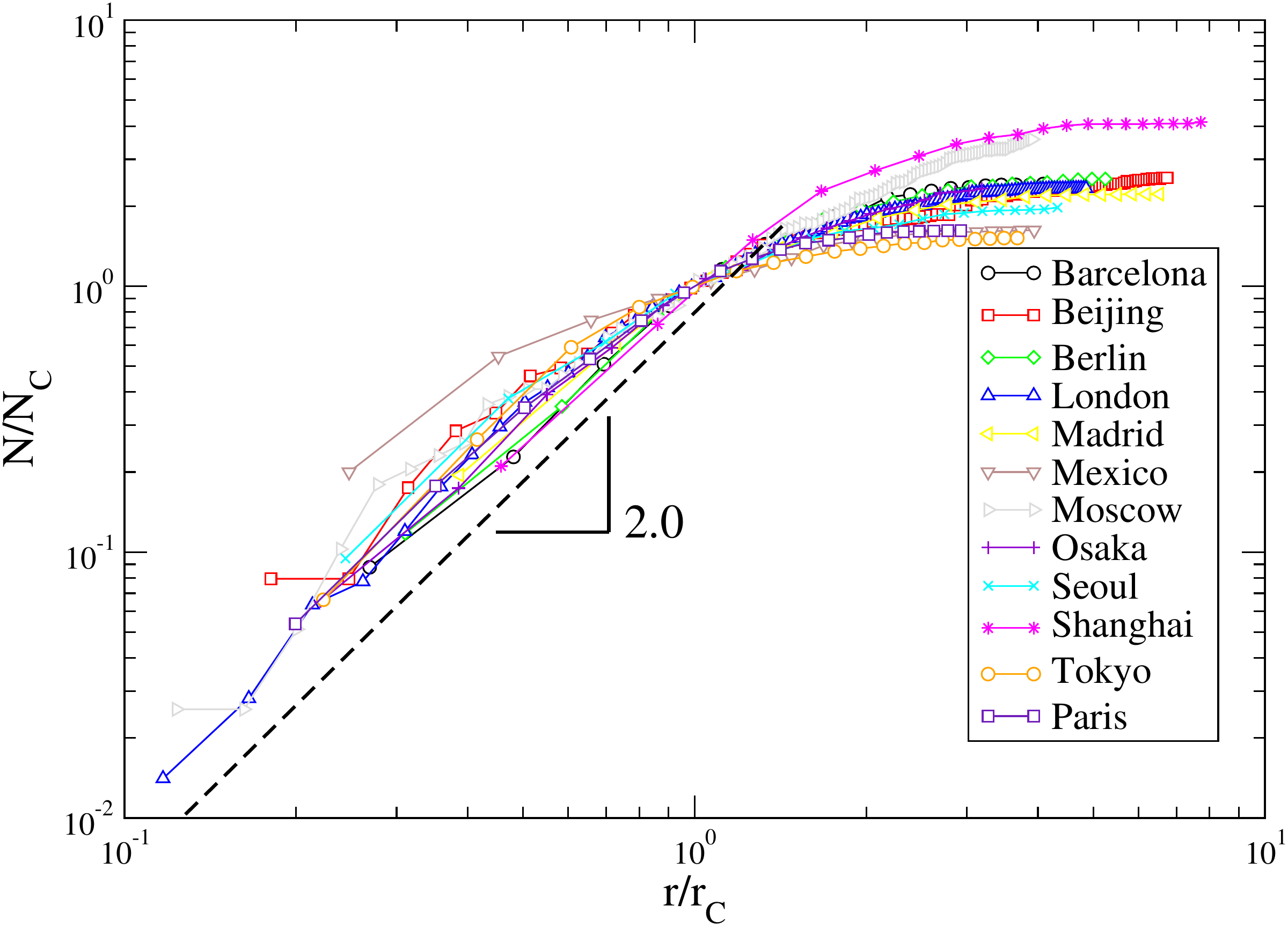}\\\\
\footnotesize(B)&\\
&
\includegraphics[width=0.7\linewidth]{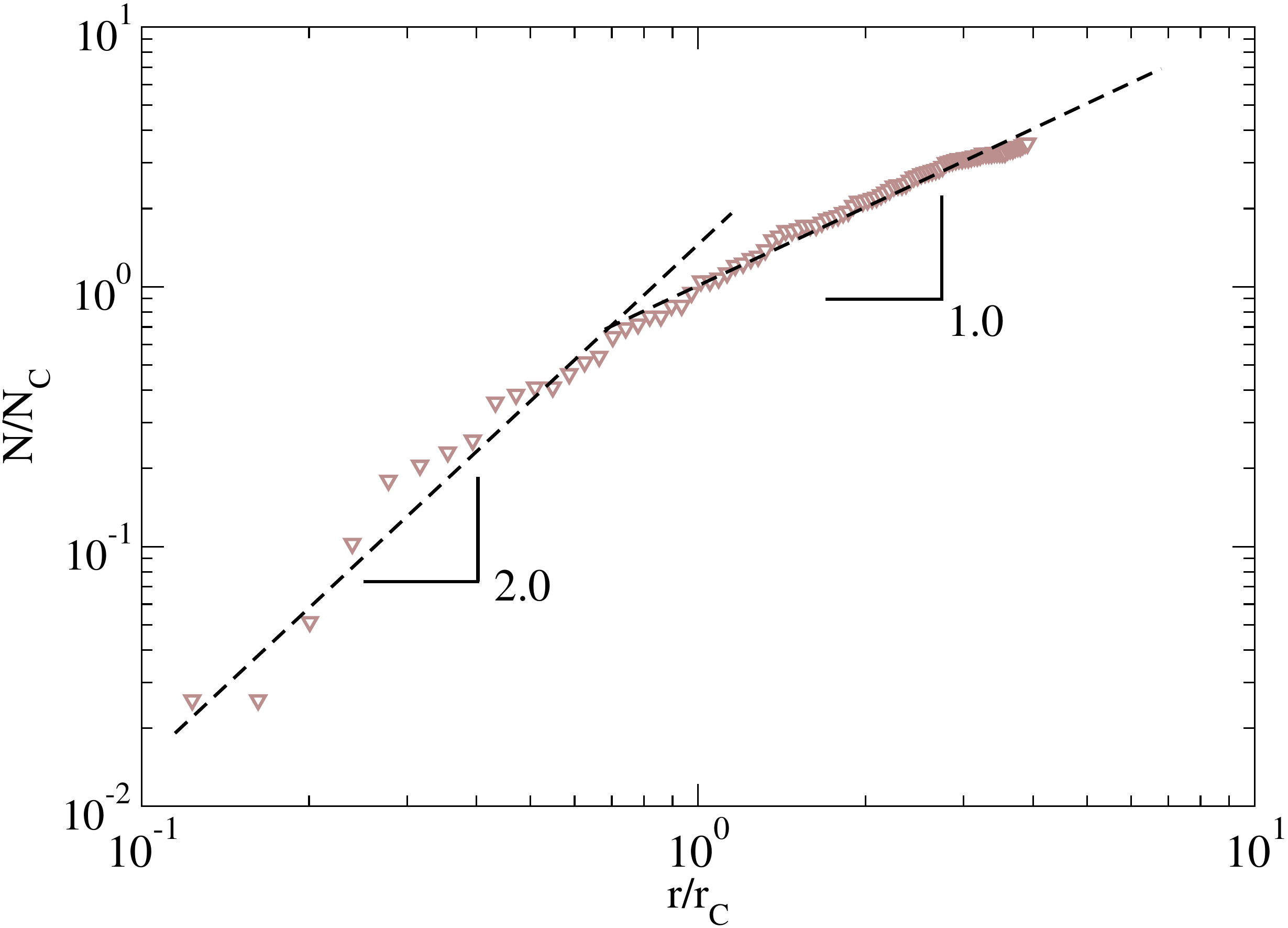}
\end{tabular}
 \caption{(A) Rescaled number of stations at distance $r$ from the
    barycenter as a function of the rescaled variable $r/r_C$ where
    $r_C$ is the size of the core defined as $N(r=r_C)=N_C$ (shown
    here in loglog). The dotted line represents a power law $\sim r^2$
    and serves as a guide to the eye. (B) Case of Moscow where the two
  regimes ($r< r_C$ and $r> r_C$) with their different exponents
  are visible (the dotted lines serve here as a guide to the eye). }\label{fig:Nvsr}
\end{figure}

Next, we can rescale $r$ by $r_C$ and $N(r)$ by $N_C$ and we then
obtain the results shown in the Fig.~\ref{fig:Nvsr}.  

This figure displays several interesting features. First, the short
distance regime $r<r_C$ is well described by a behavior of the form
$N(r)\sim \rho_C\pi r^2$ consistent with a uniform density $\rho_C$ of
core stations. For very large distances, we observe for most networks
a saturation of $N(r)$. The interesting regime is then for
intermediate distances when $r$ is larger than the core size but
smaller than the maximum branch size $r_{\text{max}}$. This intermediate
regime is characterized by different behaviors with $r$.  A similar
result was obtained earlier \cite{Benguigui:1991} where the authors
observed for Paris that $N(r>r_c)\sim r^{0.5}$, a result that was
at that time difficult to understand in the framework of fractal
geometry.

Here we show that these regimes can be understood in terms of
the core and branches model, with the additional factor that the
spacing between consecutive stations is increasing with $r$. Within
this picture (and assuming isotropy), $N(r)$ is given by
\begin{equation}
N(r)\sim 
\begin{cases}
\rho_C\pi r^2\;\;\;&\text{for}\;\;\; r< r_C\\
\rho_C\pi
r_C^2+{\cal N}_B\int_{r_C}^r\frac{dr}{\Delta(r)}\;\;\;&\text{for}\;\;\;r_C<r<r_{\text{max}}\\
N &\text{for}\;\;\;r>r_{\text{max}}\\
\end{cases}
\label{eq:scaling}
\end{equation}
where $N$ is the total number of stations, ${\cal N}_B$ is the number of
branches and $\Delta(r)$ is the average spacing between stations on
branches at distance $r$ from the barycenter. 

In order to test this shape, we can determine the
various parameters of Eq.~(\ref{eq:scaling}) --- namely ${\cal N}_B$, $N_C$, $r_C$, and $\Delta(r)$ --- and
plot the resulting shape of Eq.~(\ref{eq:scaling}) against the empirical data. It is easy to determine
empirically the numbers ${\cal N}_B$, $N_C$, and $r_C$ but the
quantity $\Delta(r)$ is extremely noisy due to the small number of
points (all these numbers are determined for the year $2009$),
especially for large values of $r$ closest to $r_{\text{max}}$, at a
distance where, often, there is no more than a handful of stations.

The less noisy situation is obtained in the case of Moscow which has
long branches and for which we obtain a interstation spacing roughly
constant. In this case we obtain for $r>r_C$ a behavior of the form
$N(r)\sim {\cal N}_Br$ (see Fig.~\ref{fig:Nvsr}b).

\begin{figure}[h]
\centering
\vspace{2em}
\begin{tabular}{c}
\includegraphics[width=0.8\linewidth]{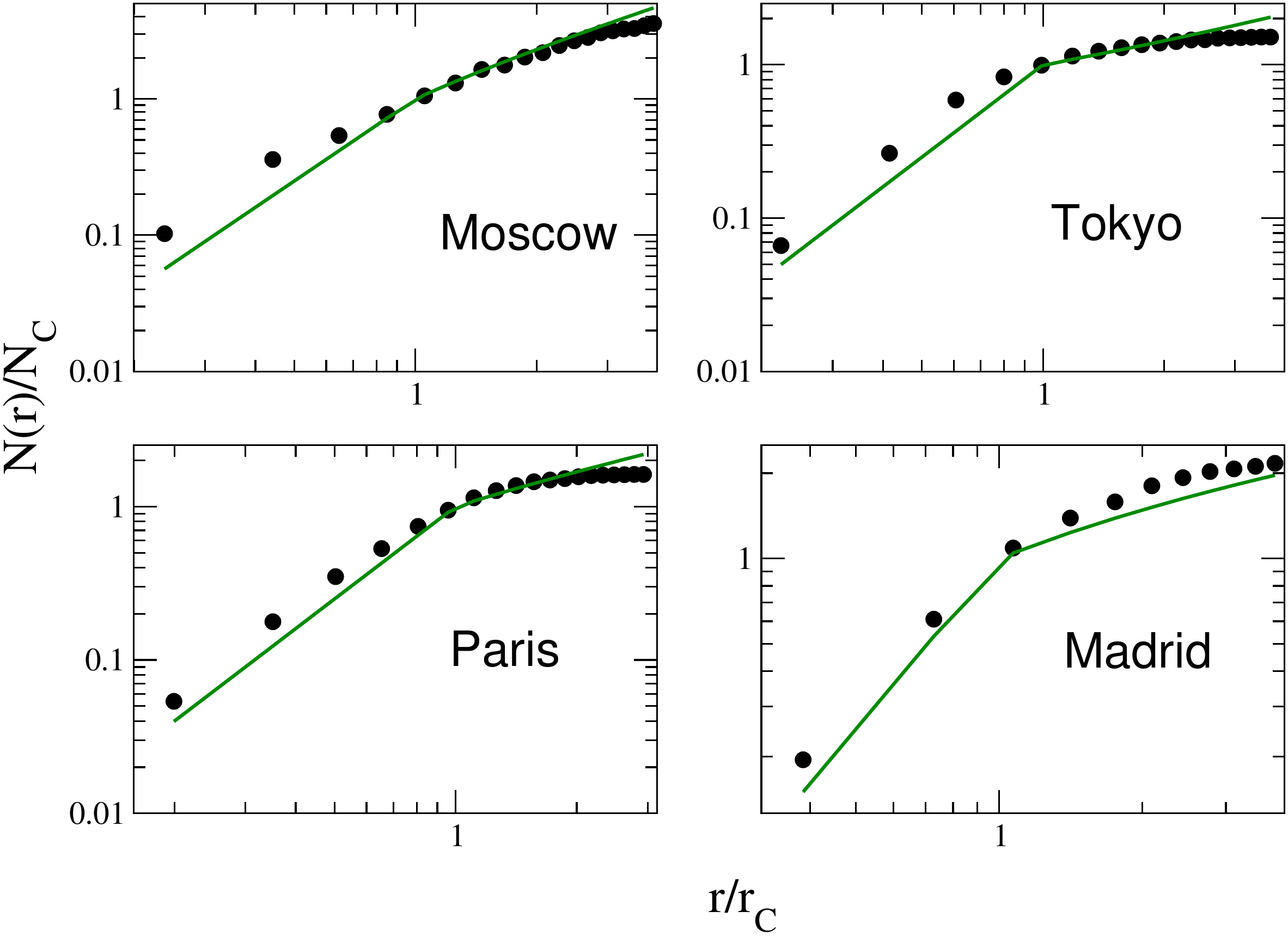}
\end{tabular}
 \caption{ $N(r)/N_C$ versus $r/r_C$ for Moscow, Tokyo, Paris, and
   Madrid (from top to bottom and left to right). The circles
   represent the data and the green solid line the fit using
   Eq.~(\ref{eq:scaling}) with parameters estimated from the empirical
   data.}\label{fig:Nfit}
\end{figure}

More generally, the large
distance behavior $r_C<r<r_{\text{max}}$ will be of the form
\begin{equation}
N(r_C<r<r_{\text{max}})\sim r^{1-\tau}
\label{eq:Nr}
\end{equation}
where $\tau$ denotes the exponent governing the interspacing decay
$\Delta(r)\sim r^{\tau}$. For most networks, the regime $r_C<
r<r_{\text{max}}$ is small and as already mentioned $\Delta(r)$ is very
noisy. Rough fits in different cases give a behavior for
Eq.~(\ref{eq:scaling}) consistent with data (see Fig.~\ref{fig:Nfit}).

In particular, for Moscow which has long branches, we observe a
behavior consistent with $\Delta(r)\simeq \text{constant}$ while for
the other networks, we observe an increasing trend but an accurate
estimate of $\tau$ is difficult to obtain, given the small variation
range of $r$ --- with no more than one decade of available data. For example, 
a fit over this decade of data gives
for Paris $\tau\approx 0.5$ (with $r^2=0.74$) in agreement with the
result obtained in \cite{Benguigui:1991}. Despite the difficulty of
obtaining accurate quantitative results, more data is needed to have a
definite answer and so far we can only claim that the data are not
inconsistent with the behavior Eq.~(\ref{eq:scaling}), which supports
our picture of a long time limit network shape made of a core and
radial branches.

\section{Discussion}
\label{sec:5}

In summary, we have observed a number of similarities between different
subway systems for the world's largest cities, despite their
geographical and historical differences.

First, we have shown that the largest subway networks exhibit a
similar temporal decrease of most fluctuations around their long term
stable values and thus converge to {a similar} structure. We
identified and characterized the shape of this long time limiting
graph as a structure made of core and branches which appears to be
relatively independent of the peculiar historical idiosyncracies
associated with the evolution of these particular cities.  

For large networks, we generally observe a fraction of branches of
about $45\%$ for most networks, and a ratio for the spatial extensions
of branches to the core of about $2$. The number of branches scales
roughly as the square root of the number of stations. The core of
these different city networks has approximately the same average
degree which is increasing with network size, from $\approx 2$ to
$\approx 2.4$ when $N\approx 100$, after which it approximately
remains within the interval $[2.3,2.5]$ (with moderate
fluctuations). The fraction of $k=2$ nodes in the core is generally
larger than $60\%$.

In addition, this picture of a core with branches and an increasing
spacing between consecutive stations on these branches is confirmed by
spatial measurements such as the number of stations at a given
distance $r$ and provides a natural interpretation to these measures.


The evolution of networks in general and urban networks in particular
represents an exciting unexplored problem which mixes spatial and
topological properties in unusual and often counterintuitive ways. They 
require a specific set of indicators that describe these phenomena. Other data such
as population density, land use activity distribution, and traffic
flows are likely to bring relevant information to this problem and
would undoubtedly enrich our study. We believe however that the
present approach represents an important exploratory step in our understanding and
is crucial for the modeling of the evolution of urban networks. In
particular, the existence of unique long-time limit topological and
spatial features is a universal signature that fundamental mechanisms,
independent of historical and geographical differences, contribute to the
evolution of these transportation networks.

{\bf Acknowledgments} We thank anonymous referees for very useful and
interesting comments.

\end{document}